\documentstyle{article}
\begin{document}
\begin{center}
{\Large \bf
Elastic scattering and bound states in the Aharonov-Bohm potential superimposed by an attractive $\rho^{-2}$ potential}
\\[0.5cm]
{\bf  J\"urgen Audretsch$^{\dagger,}$\footnote{{\it e-mail: Juergen.Audretsch@uni-konstanz.de}}, Vladimir D. Skarzhinsky$^{\dagger, \ddagger,}$\footnote{{\it e-mail: vdskarzh@sgi.lebedev.ru}} and Boris L. Voronov$^{\ddagger,}$\footnote{{\it e-mai voronov@
td.lpi.ac.ru}} }
\\[0.5cm]
$^\dagger$Fakult\"at f\"ur Physik der Universit\"at Konstanz,
Postfach M 673,\\ D-78457 Konstanz, Germany \\[0.5cm]
$^\ddagger$P.~N.~Lebedev Physical Institute, Leninsky prospect 53,
Moscow 117924, Russia

\vspace{0.5cm}

\begin{abstract}

\noindent
We consider the elastic scattering and bound states of charged quantum
particles moving in the  Aharonov-Bohm and an attractive $\rho^{-2}$ potential
in a partial wave approach. Radial solutions of the stationary Schr\"{o}dinger
equation are specified in such a way that the Hamiltonian of the problem is
self-adjoint. It is shown that they are not uniquely fixed but depend on open
parameters. The related physical consequences are discussed. The scattering
cross section is calculated and the energy spectrum of bound states is
obtained. 

\bigskip

\end{abstract}
\end{center}


{\bf 1. Introduction}

\medskip

The famous Aharonov-Bohm (AB) effect \cite{Aharonov59} has been the subject of many investigations in the past (compare \cite{Olariu85,Peshkin89}). During the last years novel applications have been discussed which open new areas of research. Detailed calculations of QED processes in the presence of the magnetic string has been carried out. For example, the cross sections of bremsstrahlung and pair production have been evaluated analytically \cite{our papers}. Recently the AB effect found an unexpected application in the domain of atomic interferometry with neutral atoms. Improving a suggestion in \cite{Wilkens94}, the following experimental set up has been proposed in \cite{Wei95}: The radial electric field $\mbox{\boldmath $E$}$ of a straight charged wire with homogenous charge density $\kappa_0$ polarizes scattered neutral atoms. A uniform magnetic field $\mbox{\boldmath $B$}$ is applied parallel to the wire. The atoms with mass $M_0$ and electric polarizability $\alpha$ moving in these two fields will 
then acquire a quantum phase. The respective Lagrangian for these atoms is
\begin{equation}\label{L}
L={1\over 2}M v^2+\alpha [\mbox{\boldmath $B\times E$}]\cdot {\mbox{\boldmath $v$}\over c}+{1\over 2}\alpha E^2
\end{equation}
with $M=M_0+\alpha B^2/c^2$ if terms of the order $v^2/c^2$ are neglected. The corresponding stationary Schr\"{o}dinger equation is in cylindrical coordinates $\rho$ and $\varphi$ (we omit the trivial $z$-dependence and put $\hbar=1$, $c=1.$)
\begin{equation}\label{Se}
\left\{{1\over 2M}\left[{\partial^2\over\partial\rho^2}
+{1\over\rho}{\partial\over\partial\rho}
-{1\over\rho^2}\left(-i {\partial\over\partial\varphi} - \beta \right)^2\right]
+{\kappa^2\over \rho^2} + {\cal E}\right\} \psi(\rho,\varphi) = 0\,,
\end{equation}
The {\sl magnetic-field parameter} $\beta$ is thereby $\beta=\alpha\kappa_0/2\pi c$ and the {\sl charge parameter} $\kappa$ is $\kappa=\alpha\kappa_0^2/8\pi^2.$ ${\cal E}=p^2/2M$ is the energy of particles with momentum $p.$ The magnetic field yields no forces on the particles, nevertheless a pure topological phase is attached to the wave function. Atom interferometer experiment to measure this effect are on their way. Results for the limiting case of a vanishing magnetic field ($B=0$) are described in \cite{Novak98}. A computer simulation has been performed in \cite{Denschlag97}. 

There is a physically different set up which leads to the same differential equation (\ref{Se}). It makes the relation to the usual AB situation evident.
The motion of a particle with mass $M$ and charge $e$ in the AB potential, 
\begin{equation}\label{AB}
\mbox{\boldmath $A(r)$} = {\Phi\over 2\pi}{[\mbox{\boldmath $n\times r$}]\over\rho^2}\,,\quad \rho^2= r^2-(\mbox{\boldmath $n\cdot r$})\,,
\end{equation}
(\mbox{\boldmath $n$} is a unit vector along the magnetic string and $\Phi$ is the magnetic flux) which is superimposed by the attractive electric potential of an electrically charged string in $\mbox{\boldmath $n$}$-direction
\begin{equation}\label{U}
U(\rho)=-{\kappa^2\over\rho^2}\,, \quad \kappa^2\geq 0
\end{equation}
is quantum mechanically described by the Hamiltonian operator\begin{equation}\label{H}
H = {1\over 2M}\left[-i\mbox{\boldmath $\nabla_r$}-e\mbox{\boldmath $ A$}(\mbox{\boldmath $r$})\right]^2 - {\kappa^2\over\rho^2}
\end{equation}
obtained by the canonical quantization procedure. Referring to cylindrical
coordinates this leads again to the Schr\"{o}dinger equation (\ref{Se}) where
the magnetic-field parameter $\beta$ agrees now with the flux
$\beta=e\Phi/2\pi.$ Below we will restrict $\beta$ to the interval
$0\leq\beta<1$ because addition of an integer number does not result in a
physical effect.

In previous papers we had introduced absorption of atoms on the surface of a
wire with finite radius \cite{Audretsch98,Audretsch99}. The same problem has
been treated in \cite{Leonhardt98}. In this paper we will take the two
singular potentials $\mbox{\boldmath $A$}$ and $U$ (which both lead to terms
proportional to $\rho^{-2})$ seriously for all values of $\rho$ including the
limit $\rho\rightarrow 0.$ That in this case severe physical problems may
arise, has already been pointed out in \cite{Hagen96}. The reason for this is
that it is not enough to consider the operator $H$ of (\ref{H}) on its own,
but the domain has to be specified on which $H$ becomes a self-adjoint and
therefore quantum mechanically admissible operator. This can be obtained by a
self-adjoint extension \cite{Reed75}. It is a well known fact that such an
extension may be unique or non-unique. Our aim is to answer this for our case
in performing the extension procedure. If it turns out to be unique, then the
self-adjoint operator with singular potentials may be regarded as a limiting
case of some well defined physical situation free of singular potentials. If
on the other hand the extension is non-unique it may be regarded as the
limiting case of many different physical set ups. In both cases it is a task
to specify the underlying non-singular situations.

%

To treat the problem we introduce a partial wave decomposition in Sect. 2.
That this is feasible for the scattering problem although the potentials in
question are far reaching, has been demonstrated \cite {Audretsch99}. In Sect.
3 we find normalizable radial functions of positive energy for the
Schr\"{o}dinger equation (\ref{Se}). They contain arbitrary coefficients
and occur to be nonorthogonal at different values of momentum $p$ unless these
coefficients are fixed appropriately. To do this we use a {\sl ``pragmatic
approach''} \cite{Audretsch95} which is based on direct evaluation of the
normalization integral. Setting the nonorthogonal terms in this
integral equal to zero we obtain the conditions under which the Hamiltonian $H$
is a self-adjoint operator on the corresponding radial functions. Being
equivalent to the standard procedure of the self-adjoint extension this
approach is more transparent since it uses the fact that any self-adjoint
operator possesses an orthogonal set of eigenfunctions. But having cured the
mathematical problem, a physical problem arises: It turns out that the
resulting complete orthogonal set of scattering states depends on an infinite
number of {\sl open} or free {\sl parameters}. What are the physical set ups
and the corresponding scattering states from which they can be obtained in an
appropriate limit?   

The cylindrically symmetric hard core model is the simplest model to think
of in which the singularity of the potentials in $\rho=0$ is avoided. We show
in Sect. 4 that the limit of vanishing radius of the hard core cylinder can
not be used to fix free parameters. So it remains an open question which
realistic physical models lead in an appropriate limit to the admissible
solutions found in the ``pragmatic approach''.  In Sect. 5 we return to the
Schr\"{o}dinger equation (\ref{Re}) and its solutions with open parameters and
discuss the general traits of the elastic scattering. We study the
differential and the total cross sections as functions of the physically
specified parameters $\beta, \kappa$ and $p.$ In Sect. 6 we turn to the bound
states and the related energy spectrum. 

In Appendix A the procedure of obtaining the self-adjoint extension is described in a mathematically more rigorous way. It is shown that the results agree with those obtained in the ``pragmatic approach''.

\bigskip

{\bf 2. Partial-wave approach to elastic scattering}

\medskip

To describe scattering we have to find a solution of Eq. (\ref{Se}) which satisfies the scattering boundary condition for $\rho\rightarrow\infty$
\begin{equation}\label{bc}
\psi (\vec{p}\,;\rho,\varphi)\rightarrow e^{i\beta (\varphi-\pi)}\,e^{ip\rho\cos\varphi} + f(\varphi)\; {e^{ip\rho}\over \sqrt{\rho}}\,,
\end{equation}
where $p := \sqrt{2M{\cal E}}$. Note that the first term, which represents the wave function $\psi_{\rm in}(\rho, \varphi)$ of the incoming flux, has the form of a modified plane wave (compare \cite{Audretsch99}). We assume that the angle $\varphi_p$ under which the particles fall in, is equal to zero. For arbitrary $\varphi_p$ replace $\varphi\rightarrow\varphi-\varphi_p.$

We assume rotational and translational symmetry and solve Eq. (\ref{Se}) using a partial wave decomposition
\begin{equation}\label{Rs}
\psi(\rho, \varphi) = \sum_{m=-\infty}^{\infty} R_m(\rho)\,e^{im\varphi}\,, \quad m=0, \pm 1, \pm 2, ...\,.
\end{equation}
$m$ is the orbital angular-momentum quantum number. That this is a well defined approach in our case although it includes Aharonov-Bohm scattering as a limiting case, has been shown in \cite{Audretsch99}. The radial functions $R_m (\rho)$ satisfy the Bessel equation
\begin{equation}\label{Re}
R''_m (\rho)+{1\over\rho}\,R'_m(\rho)-{\nu^2\over\rho^2}\,R_m(\rho)
+p^2\,R_m(\rho) = 0
\end{equation}
where we have introduced the {\sl effective quantum number} $\nu$ and the new {\sl charge parameter} $\gamma$
\begin{equation}\label{nu}
\nu^2:= (m-\beta)^2-\gamma^2\,, \quad \gamma^2 := 2 M \kappa^2\,.
\end{equation}
$\gamma\geq 0$ represents because of (\ref{U}) the strength of the attractive potential, $\nu^2$ depends apart from $\beta$ and $\gamma^2$ on the orbital quantum number $m$ (we omit for $\nu$ the index $m$). Because $\gamma$ and the magnetic-field parameter $\beta$ specify the physical situation, we call them together with the momentum $p$ the {\sl physical parameters}. If we describe with the equations above the scattering of neutral polarizable atoms by the charged wire in the uniform magnetic field \cite{Wei95}, the parameter $\nu^2$ acquires an additional term from the magnetic mass: $\gamma^2\rightarrow \gamma^2+\beta^2.$

The scattering amplitude $f(\varphi)$ can be expressed in terms of phase shifts $\delta_m$ which are defined by the asymptotic form of the radial functions
\begin{equation} \label{asymR}
R_m(\rho) \rightarrow \sqrt{1\over 2\pi p\rho} \left[e^{-i(p\rho-\pi m-{\pi\over 4})} + S_m e^{i(p\rho-{\pi\over 4})}\right], \quad S_m := e^{2i\delta_m}\,.
\end{equation}
We take into account the asymptotic behaviour of the radial functions $R_m^{\rm in}$ of the incoming wave (compare \cite{Audretsch99})
\begin{equation}\label{Rin}
\psi^{\rm in}(\rho, \varphi) = e^{i\beta (\varphi-\pi)\,e^{ip\rho\cos\varphi}} = \sum_{m=-\infty}^{\infty} R_m^{\rm in}(\rho)\,e^{im\varphi}\,,
\end{equation}
\begin{equation} \label{asymRin}
R_m^{\rm in}(\rho) \rightarrow \sqrt{1\over 2\pi p\rho} \left[e^{-i(p\rho-\pi m-{\pi\over 4})} + \cos\pi\beta e^{i(p\rho-{\pi\over 4})}\right]\,.
\end{equation}
Then the scattering amplitude reads
\begin{equation} \label{f}
f(\varphi) = {1\over\sqrt{2\pi}}\sum_{m=-\infty}^{\infty} f_m\,e^{im\varphi},
\quad f_m = {e^{-i{\pi\over 4}}\over\sqrt{p}}\left(S_m-\cos\beta\right)\,.
\end{equation}

\bigskip

{\bf 3. Self-adjoint Hamiltonian and the appearance of open parameters}

\medskip

The Hamiltonian $H$ of Eq. (\ref{Se}) contains two singular potentials, the AB-potential $A_\varphi (\rho)$ of Eq. (\ref{AB}) and the attractive scalar potential $U(\rho)$ of Eq. (\ref{U}). Both lead to terms proportional to $\rho^{-2}.$ To make $H$ a quantum mechanical operator one has to fix the domain of solutions on which it becomes a self-adjoint operator. We describe the corresponding self-adjoint extension approach in detail in the Appendix A. An alternative quick and simple procedure leading to the same results is the ``pragmatic approach'' which we have applied earlier in \cite{Audretsch95}. It is based on the demand that the radial functions $R_m (\rho)$ should be not only normalizable with regard to $\rho$ but also orthogonal for different values of $p:$
\begin{equation} \label{nc}
\int_0^{\infty} R^*_m(p'\rho)R_m(p\rho)\rho d\rho = {\delta(p-p')
\over\sqrt{pp'}}\,.
\end{equation}
This turns out to be equivalent to the restriction to solutions with specific  boundary conditions for the radial functions at $\rho=0$ which imply self-adjointness. For the pure AB potential this problem has been considered by many authors. 

It is important to note that the parameter $\nu^2$ of Eq. (\ref{nu}) can have negative values for any values of the physical parameters $\beta$ and $\gamma.$ This happens for non-vanishing charge parameter $\gamma\neq 0$ and modes with orbital quantum numbers $m$ out of the interval $(\beta-\gamma<m<\beta+\gamma).$ For the pure Aharonov-Bohm case ($\gamma=0$) there are no such modes. They are possible here because there is in addition the attractive potential $U(\rho).$ 

For the normalizable solutions we have therefore to distinguish three domains of $\nu^2$ which correspond for given physical parameters $\beta$ and $\gamma$ to different intervals of the quantum number $m.$ The conditions $\nu^2>0$ and $\mu := |\nu|> 1$ characterize the intervals of large positive and negative $m$ extending to $\pm\infty.$ (In the paper \cite{Audretsch99} we used the definition $\mu=\Im \nu$ which can differ in sign with the new one). The corresponding normalizable solutions of Eq. (\ref{Re}) are
\begin{equation} \label{rs1}
R_m(\rho) = c_m J_{\mu}(p\rho), \quad |m-\beta|> \sqrt{1+\gamma^2}\,, \quad |c_m|=1\,.
\end{equation}

The normalizable solutions with $\nu^2>0$ and $\mu<1$ have the form
\begin{equation} \label{rs2}
R_m(\rho) = a_m J_{-\mu}(p\rho) + b_m J_{\mu}(p\rho), \quad \gamma<|m-\beta|< \sqrt{1+\gamma^2}
\end{equation}
with arbitrary coefficients $a_m$ and $b_m.$ If $\gamma\neq 0$ the condition in Eq. (\ref{rs2}) can be fulfilled for not more than for two values of $m,$ depending on the values of $\beta$ and $\gamma.$ For the pure AB case ($\gamma=0$), for example, we can have only $m=0$ and $m=1.$ This domain is with respect to the values of $m$ an intermediate domain.

Turning to $\nu^2<0$ what is possible for $\gamma\neq 0$ we find
\begin{equation} \label{rs3}
R_m(\rho) = a_m J_{-i\mu}(p\rho) + b_m J_{i\mu}(p\rho), \quad |m-\beta|< \gamma\,.
\end{equation}
These solutions with the smallest values of $m$ have to be taken into account whenever the scalar potential is large enough so that $\gamma > \min [\beta, 1-\beta].$

The solutions (\ref{rs1}) are orthogonal for different $p.$ This is in general not the case for the solutions (\ref{rs2}) and (\ref{rs3}) if the  coefficients $a_m$ or $b_m$ are not specified further. Because of the singular potentials, the Hamiltonian of Eq. (\ref{Re}) is not self-adjoint if applied to the domain of all possible solutions Eqs. (\ref{rs1}) - (\ref{rs3}). Following the ``pragmatic approach'' to self-adjointness we will now impose the condition (\ref{nc}). Using the relations (see Appendix B)
\begin{equation} \label{int1}
\int_0^{\infty} J_\nu (p'\rho) J_\nu(p\rho)\rho d\rho = {\delta(p-p')
\over\sqrt{pp'}}
\end{equation}
and
\begin{equation} \label{int2}
\int_0^{\infty} J_{-\nu}(p'\rho) J_{\nu}(p\rho)\rho d\rho =
{\delta(p-p')\over\sqrt{pp'}} \cos\pi\nu
+{2\sin\pi\nu\over\pi (p^2-p'^2)}\left(p\over p'\right)^\nu \,,
\end{equation}
we obtain from Eq. (\ref{rs2}) for the intermediate interval of quantum numbers $m$:
\begin{eqnarray} \label{nc1}
&&\int_0^{\infty} R^*_m(p'\rho)R_m(p\rho)\rho d\rho
= [a_m^*a_m+b_m^*b_m+\cos\pi\mu(a_m^*b_m+b_m^*a_m)]
{\delta(p-p')\over\sqrt{pp'}} \nonumber \\
&+& {2i\sin\pi\mu\over\pi(p^2-p'^2)}
\left[\left({p\over p'}\right)^{\mu}a_m^*(p')b_m(p)-
\left(p\over p'\right)^{-\mu}b_m^*(p')a_m(p)\right] \,.
\end{eqnarray}
It is the second term which may prevent orthogonality. To establish orthogonality we have to demand
\begin{equation}\label{ba2}
{b_m^*(p')\over a_m^*(p')}p'^{2\mu}={b_m(p)\over a_m(p)}p^{2\mu}\,,
\end{equation}
what implies that the expressions on both side of this equation are independent of $p$ and real. Accordingly we have for $\nu^2>0$ and $\mu<1$
\begin{equation} \label{rs2'}
R_m(\rho) = c_m \left[\lambda_m \left(p\over M\right)^{2\mu} J_{-\mu}(p\rho) +  J_{\mu}(p\rho)\right]\,, \quad \gamma<|m-\beta|< \sqrt{1+\gamma^2}
\end{equation}
with dimensionless real parameters $\lambda_m$ fulfilling $|c_m|^2 \left[\lambda_m^2 \left(p\over M\right)^{4\mu} + 2\lambda_m\left(p\over M\right)^{2\mu} \cos\pi\mu +1\right]=1$ which are not specified by the pragmatic approach. We will confirm this more rigorously in Appendix A. The procedure to establish self-adjointness does therefore not lead to a unique solution for given $m$ but to a whole set of solutions parametrized by the set of open parameters \{$\lambda_m$\}.

Turning to quantum numbers $m$ with  $\nu^2<0$ we obtain with Eq. (\ref{rs3})
\begin{eqnarray} \label{nc2}
&&\int_0^{\infty} R^*_m(p'\rho)R_m(p\rho)\rho d\rho
= [a_m^*b_m+b_m^*a_m+\cosh\pi\mu(a_m^*a_m+b_m^*b_m)]
{\delta(p-p')\over\sqrt{pp'}} \nonumber \\
&-& {2i\sinh\pi\mu\over\pi(p^2-p'^2)}
\left[\left({p\over p'}\right)^{-i\mu}a_m^*(p')a_m(p)-
\left(p\over p'\right)^{i\mu}b_m^*(p')b_m(p)\right] \,.
\end{eqnarray}
For orthogonality the second term has to vanish. This amounts to
\begin{equation}\label{ba3}
{b_m^*(p')\over a_m^*(p')}p'^{-2i\mu}={a_m(p)\over b_m(p)}p^{-2i\mu}\,. \end{equation}
Accordingly the expressions on either side of this equation must be independent of $p$ and their complex conjugate must be equal to their inverse. The consequence is that the radial solutions $R_m$ of Eq. (\ref{rs3}) must be restricted to
\begin{equation} \label{rs3'}
R_m(\rho) = c_m \left[e^{i[\theta_m+2\mu\ln (p/M)]}J_{-i\mu}(p\rho) +  J_{i\mu}(p\rho)\right]\,, \quad |m-\beta|< \gamma
\end{equation}
with open dimensionless parameters $\theta_m$ out of the interval $0\leq\theta_m<2\pi$ 
and the normalization condition amounts to $2|c_m|^2[\cosh\pi\mu+\cos(2\mu\ln p/M +\theta_m)]=1. $ Again self-adjointness does not leads to a unique solution for a given m (compare App. A).

It is easily to check that the partial radial currents
\begin{equation} \label{prc}
j_m (\rho) = {p\over 2iM}[R_m^*(p\rho) R'_m(p\rho) - R_m(p\rho) R'^*_m(p\rho)]
\end{equation}
vanish for the solutions (\ref{rs1}), (\ref{rs2'}) and (\ref{rs3'}).  The
same is true for the total current.The reason for this that ingoing and
outgoing currents compensate each other, thereby reflecting the fact that the
scattering is purely elastic.

\bigskip

{\bf 4.The cylindrical hard core model}

\medskip

We have shown that for the elastic scattering of particles in the two singular
potentials the relevant Hamiltonian is self-adjoint and therefore
quantum-mechanically admissible not only for one but for a total set of
solutions parametrized by open parameters. Because there is not a unique
solution, the differential equation (\ref{Re}) can not be attributed to one
specific physical situation. The reason may be that many different physical
situations with non-singular potentials and certain boundary conditions lead
in some physical limit to the same differential equation (\ref{Re}) with two
singular potentials. One could imagine that as far as the wave functions are
concerned, each of these situations lead to one particular solution $R_m
(\rho)$ with well specified parameters $\lambda_m$ and $\theta_m.$ In the
following we want to examine if in this way the physically simplest well
defined physical situation is related to particular values of the parameters.
For this we study the model with a hard core with surface $\rho_0$ so that no
particles can penetrate the surface. This amounts to $R_m (\rho)=0$ for
$\rho_0\leq 0$ (compare \cite{Audretsch98}). The limit to be discussed is
$\rho_0\rightarrow 0.$

For quantum numbers $m$ with $\nu^2>0$ the hard core solutions
\begin{eqnarray} \label{rs0}
R^{\rm hc}_m (\rho) &=& \tilde{c}_m [J_{\mu}(p\rho_0) J_{-\mu}(p\rho) - J_{-\mu}(p\rho_0) J_{\mu}(p\rho)] \nonumber \\
&&\sim -[J_{\mu}(p\rho_0) /J_{-\mu}(p\rho_0)] J_{-\mu}(p\rho) +  J_{\mu}(p\rho)\,, \quad |m-\beta|>\gamma
\end{eqnarray}
do not contain open parameters because there are no singular potentials furthermore. Because of $\lim_{\rho_0\rightarrow 0} J_\mu (p\rho_0)/ J_{-\mu}(p\rho_0) =0$ this leads for $\rho_0\rightarrow 0$ to the solution (\ref{rs1}) for $\mu>1$  and to the solution (\ref{rs2'}) with fixed parameter $\lambda_m=0$ in the intermediate region with $\mu<1.$

For the domain of quantum numbers $m$ with $\nu^2<0$ the parameter free hard core solutions are
\begin{equation} \label{rs00}
R^{\rm hc}_m (\rho) = \tilde{c}_m [J_{-i\mu}(p\rho_0) J_{i\mu}(p\rho) - J_{i\mu}(p\rho_0) J_{-i\mu}(p\rho)] \,, \quad |m-\beta|<\gamma\,.
\end{equation}
In order to see to which values of the parameters $\theta_m$ this is related, we go to the limit $\rho_0\rightarrow 0$ and find
\begin{equation}\label{rho0}
e^{i(\theta_m+2\mu\ln p/M)} = \lim_{\rho_0\rightarrow 0} \left[-{J_{i\mu}(p\rho_0)\over J_{-i\mu}(p\rho_0)}\right]\,,\quad \ \theta_m =\lim_{\rho_0\rightarrow 0}2\mu\ln{M\rho_0\over 2}+ \pi-2\xi_m
\end{equation}
with $\xi_m$:= $\arg\Gamma(1+i\mu).$ For $\rho_0\rightarrow 0$ we have
$\theta_m\rightarrow -\infty.$ Therefore this limit of the hard core model
does not fix the open parameters $\theta_m.$ 

This is not surprising. The problem in question as it is specified by Eq.
(\ref{Se})  is explicitly scale invariant.  On the another hand, by the
appearance of open parameters  caused by the self-adjoint extension, scales
are introduced (compare the example in Sect. 6). To obtain a physical
interpretation of the open parameters from more physical models in going to a
certain limit, these models must contain scale parameters which do not
dissapear in this limit. The hard core model contains the parameter $\rho_0,$
but it dissappears in the limit $\rho_0\rightarrow 0.$ 

Taken these results together we have shown that the hard core model in the
limit of vanishing $\rho_0$ does not leads to a solution $\psi (\rho,
\varphi)$ based on the radial solutions $R_m(\rho)$ above if values of $m$ out
of the interval $|m-\beta|<\gamma$ have to be taken into account. For large
values of the charge parameter $\gamma$ this will be the case.

\bigskip

{\bf 5. Scattering amplitude and cross section}

\medskip

To work out the scattering cross section in the general case of Eqs. (\ref{Se}) and (\ref{Re}) we refer to the limit of the hard core model and take $\lambda_m=0.$ In this case there is one type of solutions for the whole quantum number domain $\nu^2>0$ so we have
\begin{equation}\label{swf}
 R_m = \left\{
\begin{array}{ll}
c_m J_\mu(p\rho) & {\rm for}\; \nu^2>0\,, \\ \\
c_m [\left[e^{i[\theta_m+2\mu\ln (p/M)]}J_{-i\mu}(p\rho) +  J_{i\mu}(p\rho)\right] &{\rm for}\; \nu^2<0\,.
\end{array}\right.
\end{equation}
Based on the scattering boundary condition (\ref{bc}) and the asymptotic form (\ref{asymRin}) of the partial waves of the distorted ingoing plane wave, we find for the coefficients
\begin{equation}\label{swfc}
c_m = \left\{
\begin{array}{ll}
e^{i\pi(m-{\mu\over 2})} & {\rm for}\; \nu^2>0\,, \\ \\
e^{i\pi m}\left[e^{-{\pi\over 2}\mu}+e^{{\pi\over 2}\mu}e^{i(\theta_m+2\mu\ln p/M)}\right]^{-1} \;, &{\rm for}\; \nu^2<0 \,.
\end{array}\right.
\end{equation}
The partial scattering amplitudes of Eq. (\ref{f}) then turn out to be
\begin{equation}\label{swfs}
S_m = \left\{
\begin{array}{ll}
e^{i\pi(m-\mu)}\;, & {\rm for}\; \nu^2>0 \\
e^{i\pi m-i(\theta_m+2\mu\ln p/M)} {e^{\pi\mu}+e^{i(\theta_m+2\mu\ln p/M)}\over e^{\pi\mu}+e^{-i(\theta_m+2\mu\ln p/M)}} \;, &{\rm for}\; \nu^2<0 \,.
\end{array}\right.
\end{equation}
Eq. (\ref{Re}) is explicitly scale-invariant because the parameter $\gamma$ and, consequently, the parameter $\nu$ are dimensionless. One could therefore expect that the phase functions are independent of the momentum $p$ and the related energy \cite{Jackiw72}. This is the case if the operator of Eq. (\ref{Re}) is essentially self-adjoint. If not, the procedure of self-adjoint extension may lead to a $p$-dependence. Eq. (\ref{swfs}) shows this explicitly for $\nu^2<0.$ It is then convenient to replace $\theta_m$ by dimensional open parameters $p_{m, 0}$ out of the interval $M e^{-\pi/\mu}\leq p_{m, 0}<M$ according to
\begin{equation}\label{p}
\theta_m=:2\mu\ln{M\over p_{m, 0}}\,,
\end{equation}
so that $S_m$ of Eq. (\ref{swfs}) depend on $p/p_{m, 0}.$ We will see below that because of the necessity for a self-adjoint extension for bound states, the energy spectrum will depend on $p/p_{m, 0}$ too, so that an arbitrary energy scale appears.  

To work out the scattering amplitude $f(\varphi)$ it is useful to separate the contribution of the AB potential in order to get convergent series (compare \cite{Audretsch99}). This leads to
\begin{eqnarray} \label{f-AB}
&&f(\varphi)={e^{-i{\pi\over 4}}\over\sqrt{2\pi p}}\sum_{m=-\infty}^{\infty} \left(S_m-\cos\beta\right) \,e^{im\varphi} \\
&&=f_{\rm AB}(\varphi)+{e^{-{i\pi\over 4}}\over\sqrt{2\pi p}} \sum_{m=-\infty}^{\infty}\left(S_m-S_m^{\rm AB}\right) \,e^{im\varphi} := f_{\rm AB}(\varphi) + {e^{-i{\pi\over 4}}\over\sqrt{2\pi p}}\,\delta\Sigma (\varphi) \nonumber
\end{eqnarray}
with
\begin{equation} \label{fAB}
S_m^{\rm AB} = e^{i\pi(m-|m-\beta|)}\,,\quad f_{\rm AB}(\varphi)= - {1\over \sqrt{2\pi p}}\,e^{-i \pi/4}\;e^{i \varphi/2}\; {\sin\pi\beta\over\sin{\varphi\over 2}}\,.
\end{equation}
For the subsequent numerical evaluation we split $\delta\Sigma (\varphi)$ into three parts,
\begin{eqnarray} \label{sigma}
\delta\Sigma (\varphi) &=& \sum_{m>\gamma+\beta}^{\infty}\left(S_m-S_m^{\rm AB}\right) \,e^{im\varphi} + \sum_{m>\gamma-\beta}^{\infty}\left(S_{-m}-S_{-m}^{\rm AB}\right) \,e^{-im\varphi} \nonumber \\
&+& \sum_{m>\beta-\gamma}^{m<\beta+\gamma}\left(S_m-S_m^{\rm AB}\right) \,e^{im\varphi}
\end{eqnarray}

The differential cross section $d\sigma/d\varphi$ is a function of the scattering angle $\varphi$ and depends on the physical parameters $\beta$ and $\gamma$ and on momentum $p$ which specify the underlying physical situation. The open parameters $\theta_m$ in Eq. (\ref{swfs}) are by definition functions of $M/p_{m, 0}$ (compare Eq. (\ref{p})). In the following we chose $\theta_m$ to be equal for all $m$ and correspondingly $p_{m, 0} := p_0.$ Equations (\ref{swfs}) and (\ref{f-AB}) then 
show that the dimensionless differential scattering cross section is a function of $\varphi$ and of the combination $p/p_0$
\begin{equation} \label{dcs}
2\pi p\,{d\sigma\over d\varphi} = \sqrt{2\pi p}\,|f(\varphi)|^2 := \eta\, (\varphi, p/p_0)\,.
\end{equation}
In Fig. 1 we show the function $\eta\,(\varphi, p/p_0)$ obtained by a numerical calculation for $p/p_0=1$ and $p/p_0=80$ (solid lines) and values of the physical parameters $\beta=0.5$ and $\gamma=9.9$. The dashed line represent the AB limit $\gamma=0.$ For small scattering angles the AB contribution dominates the effect due to the factor $\sin^2{\varphi\over 2}$ in the denominator of $f_{\rm AB}(\varphi).$  It can be seen that $\eta$ for small angles $\varphi$ is independent of $p/p_0$ and therefore independent of the open parameter $p_0.$ For larger scattering angles the AB contribution becomes negligibly small. We notice here oscillations as functions of $\varphi$ as well as a weak dependence on $p/p_0.$

\begin{figure}[h]
\let\picnaturalsize=N
\def\picsize{2.5in}
\ifx\nopictures Y\else{\ifx\epsfloaded Y\else\input epsf \fi
\let\epsfloaded=Y
{\hspace*{\fill}
\parbox{2.5in}{\ifx\picnaturalsize N\epsfxsize \picsize\fi \epsfbox{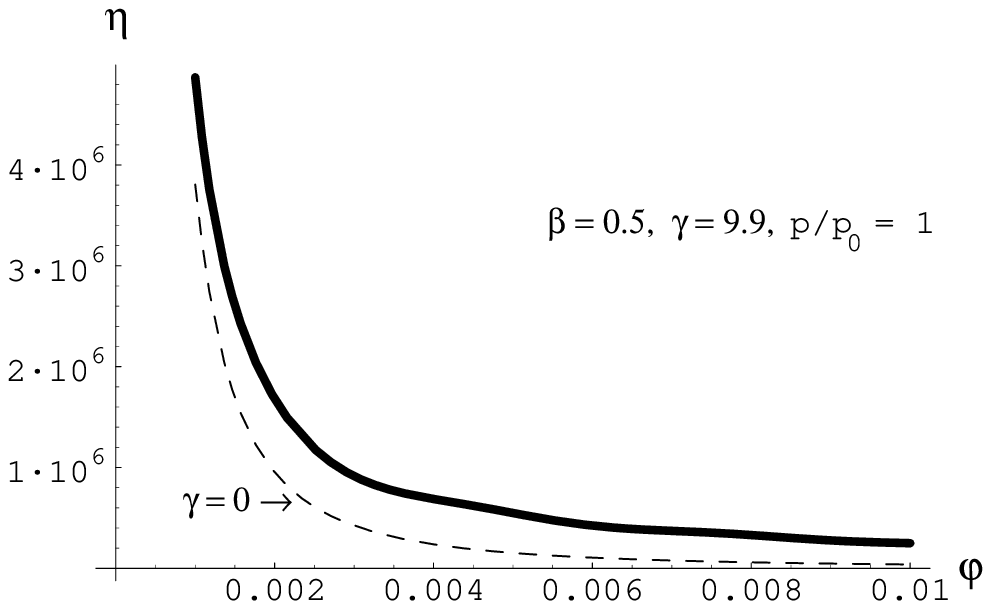}}\hfill
\hspace{1cm}
\parbox{2.5in}{\ifx\picnaturalsize N\epsfxsize \picsize\fi \epsfbox{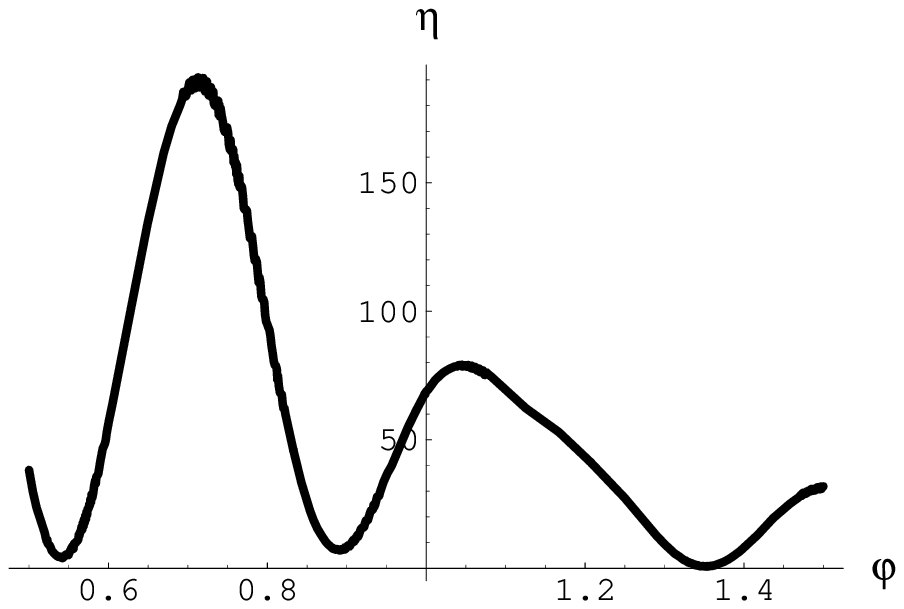}}\hfill
\hspace{1cm}
\parbox{2.5in}{\ifx\picnaturalsize N\epsfxsize \picsize\fi \epsfbox{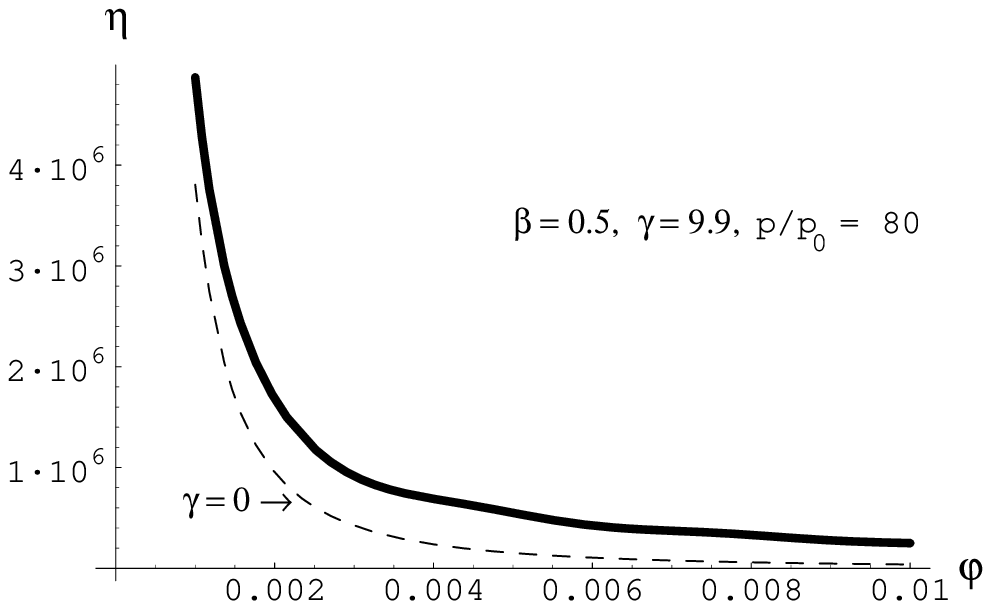}}\hfill
\hspace{1cm}
\parbox{2.5in}{\ifx\picnaturalsize N\epsfxsize \picsize\fi \epsfbox{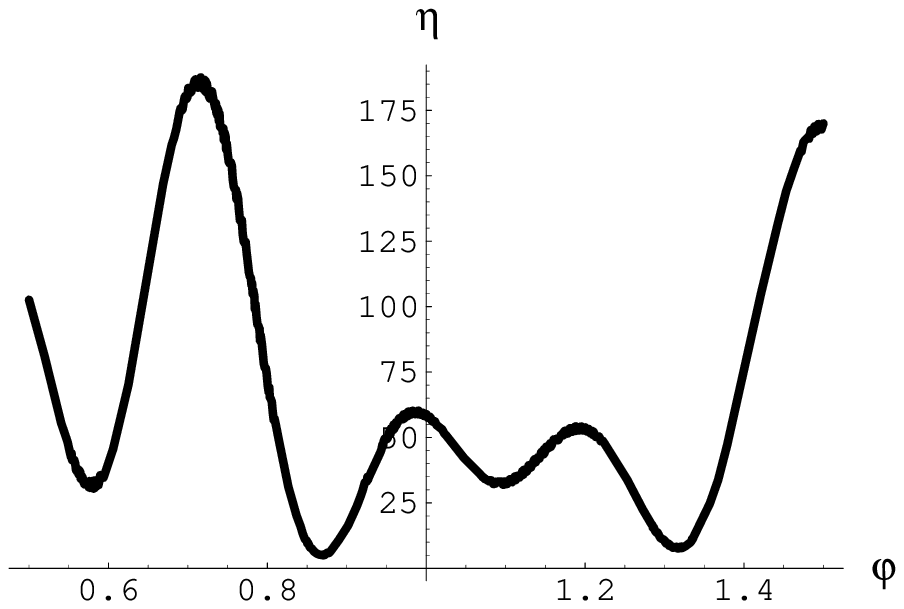}}\hspace*{\fill}}}\fi\\
\caption{Dimensionless differential cross section $\eta=2\pi p\,d\sigma/ d\varphi$ of elastic scattering as function of the scattering angle $\varphi$ for two domains of $\varphi.$ The AB cross section ($\gamma=0$) is shown as dashed line.}
\end{figure}

We turn now to the total elastic scattering cross section $\sigma.$ It is divergent in the presence of the AB potential which decreases too slowly at infinity. Therefore we work out $\sigma$ for the special case of vanishing flux parameter $\beta=0.$ Accordingly it depends only on the momentum $p,$ the charge parameter $\gamma$ and on the open parameters $p_{m, 0}.$ We obtain
\begin{equation} \label{tcs}
\sigma=\int_0^{2\pi} |f(\varphi)|^2 d\varphi = \sum_m \sigma_m =\sum_m f^*_m f_m = {2\over p} \sum_m (1-\Re S_m)
\end{equation}
with
\begin{equation}\label{res}
1-\Re S_m = \left\{
\begin{array}{ll}
1-\cos\pi m \cos\pi\mu\;, & {\rm for}\; \nu^2>0 \\
1-\cos\pi m \cosh\pi\mu+{\sinh^2\pi\mu \cos\pi m\over \cosh\pi\mu+\cos (\theta+2\mu\ln p/M)} \;, &{\rm for}\; \nu^2<0 \,,
\end{array}\right.
\end{equation}
$\theta_m$ of Eq. (\ref{p}) and $\mu=|\nu|$ of Eq. (\ref{nu}). Decomposing according to
\begin{equation}\label{dec}
\sigma = \sigma_{\nu^2>0} + \sigma_{\nu^2<0}\,,
\end{equation}
we see that $\sigma_{\nu^2>0}$ has a simple $p$-dependence of the form
\begin{equation}\label{s+}
\sigma_{\nu^2>0} = {2 \Sigma_1(\gamma)\over p}\,.
\end{equation}
where $\Sigma_1 (\gamma)$ is an increasing function of $\gamma.$ $\Sigma_1 (\gamma)$ is close to $\pi^4\gamma^4/24$ for small $\gamma$, whereas it approaches asymptotically the parabola $\pi^2\gamma^2/2$ for large $\gamma,$ see Fig. 2. 
\begin{figure}[h]
\let\picnaturalsize=N
\def\picsize{2.5in}
\ifx\nopictures Y\else{\ifx\epsfloaded Y\else\input epsf \fi
\let\epsfloaded=Y
{\hspace*{\fill}
\parbox{2.5in}{\ifx\picnaturalsize N\epsfxsize \picsize\fi \epsfbox{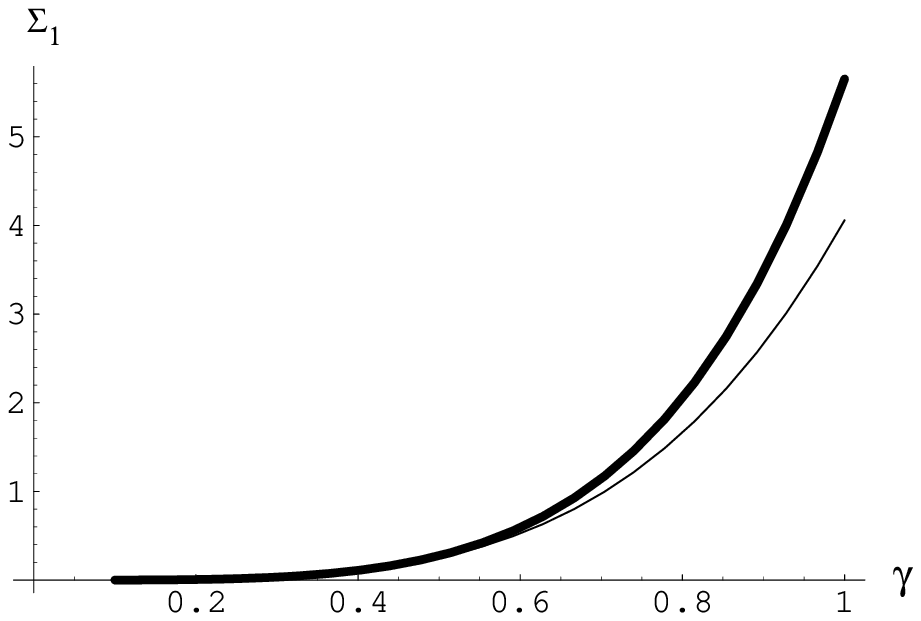}}\hfill
\parbox{2.5in}{\ifx\picnaturalsize N\epsfxsize \picsize\fi \epsfbox{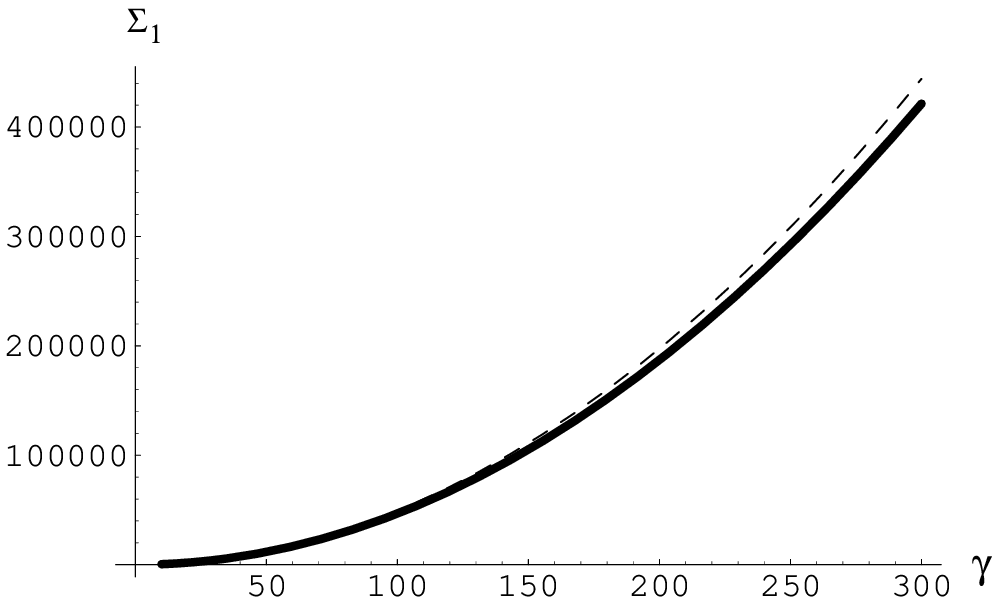}}\hspace*{\fill}}}\fi\\
\par\vspace{0.5cm}
\caption{The function $\Sigma_1 (\gamma)$ of Eq. (\ref{s+}) (bold line).The thin and dashed lines show the functions $\pi^4\gamma^4/24$ and $\pi^2\gamma^2/2$ correspondingly.}
\end{figure}

$\sigma_{\nu^2<0}$ has in contrast to $\sigma_{\nu^2>0}$ no simple dependence on the momentum $p.$ Writing 
\begin{equation}\label{s-}
\sigma_{\nu^2<0} = {2\over p} \Sigma_2(\gamma, p)\,,
\end{equation}
we can discuss the $p$-dependence of $\Sigma_2$ for a given value of the charge parameter $\gamma.$ Restricting to the domain $0<\gamma<1$ of small $\gamma,$ it is only the $s$-wave which contributes to $\sigma_{\nu^2<0}.$ Fitting the remaining open parameter as $p_{0, 0}=10$ we obtain for $\gamma=0.9$ the curves of Fig. 3. They show an oscillatory dependence on $p$ which becomes infinitely fast as $p\rightarrow 0.$ The contributions to $\sigma_{\nu^2<0}$ with $m\neq 0$ which appear for larger values of $\gamma$ show the same behaviour.

\begin{figure}[h]
\let\picnaturalsize=N
\def\picsize{2.5in}
\ifx\nopictures Y\else{\ifx\epsfloaded Y\else\input epsf \fi
\let\epsfloaded=Y
{\hspace*{\fill}
\parbox{2.5in}{\ifx\picnaturalsize N\epsfxsize \picsize\fi \epsfbox{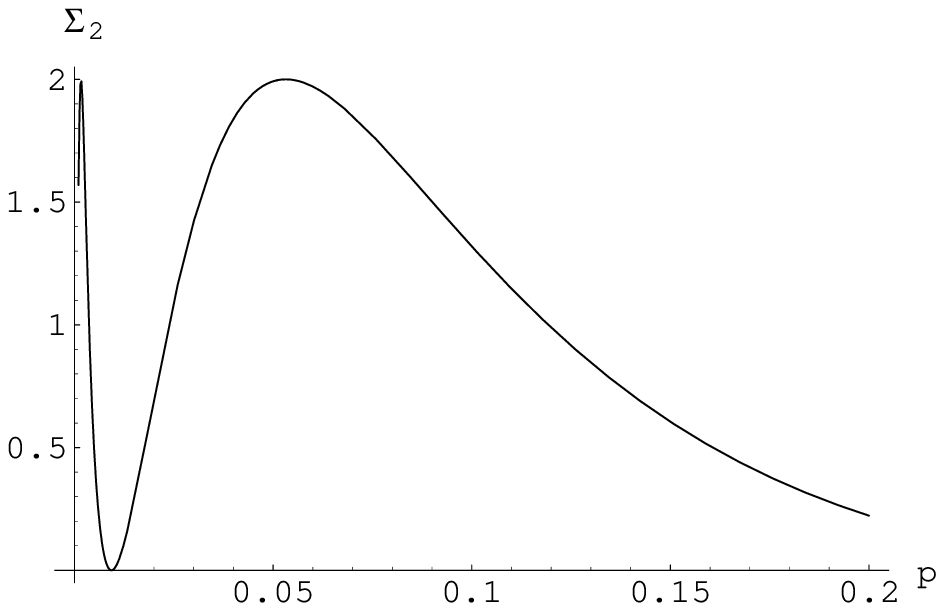}}\hfill
\parbox{2.5in}{\ifx\picnaturalsize N\epsfxsize \picsize\fi \epsfbox{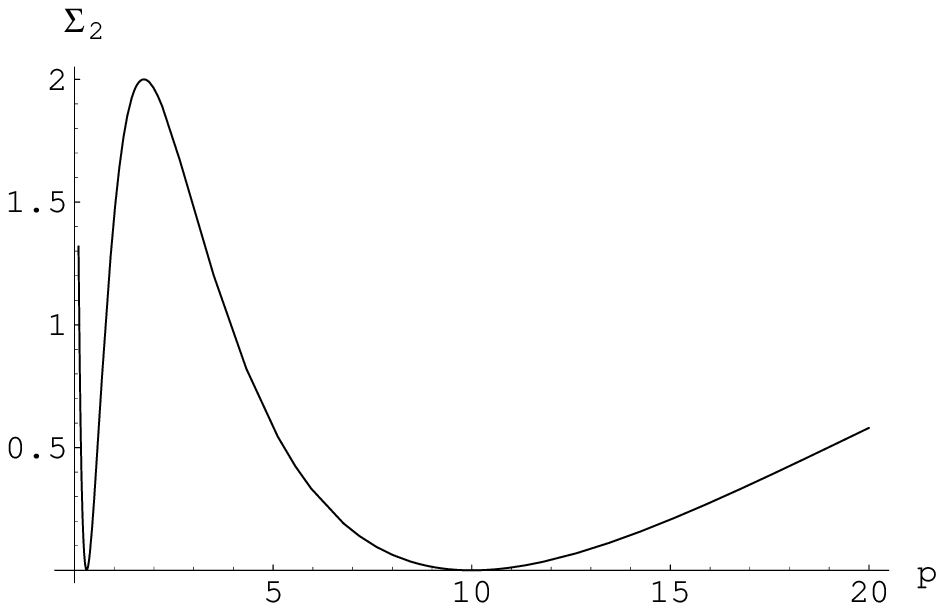}}\hspace*{\fill}}}\fi\\
\par\vspace{0.5cm}
\caption{$\Sigma_2 (p)$ of Eq. (\ref{s-}) for the parameter choice $\beta=0, \gamma=0.9$ and $p_{0, 0}=10.$ The dimensionless total cross section $0.5\,p\,\sigma(p)$ is obtained in adding to $\Sigma_2$ the constant $\Sigma_1\approx 3.48$ thus shifting up 
the curves.}
\end{figure}

\bigskip

{\bf 6. Bound states}

\medskip

Bound states in the attractive potential $U(x) = - \kappa^2/x^2$ have been considered long ago in \cite{Case50}. We generalize these results by including the AB potential $(\beta\neq 0).$ Bound states can only be found for values of $m$ with $\nu^2<0.$ The energy ${\cal E}$ is negative so that we have $p^2=-2 M {\cal E}.$ The corresponding radial equation
\begin{equation}\label{-Re}
R''_m (\rho)+{1\over\rho}\,R'_m(\rho)+{\mu^2\over\rho^2}\,R_m(\rho)
-p^2\,R_m(\rho) = 0
\end{equation}
has the general solution
\begin{equation}\label{bse}
R_m(\rho) = a_m I_{-i\mu}(p\rho)+b_m I_{i\mu}(p\rho)\,.
\end{equation}
Because the wave functions of bound states have to decrease for $\rho\rightarrow\infty$, this leads to a real $R_m (\rho)$
\begin{equation}\label{bss}
R_m(\rho) = i c_m [ I_{i\mu}(p\rho) - I_{-i\mu}(p\rho)] = {2 c_m\over\pi}\sinh \pi\mu\,K_{i\mu}(p\rho)\,,
\end{equation}
with $K_{\nu}(z)$ being the Bessel function of imaginary argument. For $\rho\rightarrow 0$ we find
\begin{eqnarray}\label{0}
R_m(\rho) &\sim& i c_m \left[\left({p\rho\over 2}\right)^{i\mu} {1\over\Gamma(1+i\mu)} - \left({p\rho\over 2}\right)^{-i\mu} {1\over\Gamma(1-i\mu)}\right] \nonumber \\
&=& {i c_m \over\sqrt{\Gamma(1+i\mu)\Gamma(1-i\mu)}} \left[e^{i\mu\ln(p\rho/2) -i\xi} - e^{-i\mu\ln(p\rho/2) +i\xi}\right] = \nonumber \\
&-&{2c_m \sinh\pi\mu\over\pi\mu}\sin\left(\mu\ln{p\rho\over 2}-\xi\right)
\end{eqnarray}
where $\xi:=\arg\Gamma(1+i\mu).$ To obtain self-adjointness we demand again orthogonality of the wave functions for different values of $p:$
\begin{eqnarray}\label{oc}
&&(p^2-p'^2)\int_0^\infty \rho d\rho R_m(p'\rho) R_m(p\rho) = \int_0^\infty {d\over d\rho}\left[ R_m(p'\rho) R'_m(p\rho) -  R_m(p\rho) R'_m(p'\rho)\right] \nonumber \\
&&= \rho \left[R_m(p\rho) R'_m(p'\rho) -  R_m(p'\rho) R'_m(p\rho) \right]\mid_{\rho=0} =0\,.
\end{eqnarray}
This condition will be fulfilled if we have for given $m$
$$
\rho \left[R_m(p\rho) R'_m(p'\rho) -  R_m(p'\rho) R'_m(p\rho)\right]\mid_{\rho=0} = {4 c_m^2 \sinh^2\pi\mu\over\pi^2\mu}\sin\left(\mu\ln {p\over p'}\right) = 0\,,
$$
or
\begin{equation}\label{s}
{p'\over p} = e^{\pi n\over\mu}, \quad \mbox{i.e.} \quad p_{m, n} = p_{m, 0}\, e^{-{\pi n\over\mu}},\quad {\cal E}_{m, n} = {\cal E}_{m, 0}\,e^{-{2\pi n\over\mu}}\,,\quad {\cal E}_{m, 0} =-{p^2_{m, 0}\over 2M} .
\end{equation}
with $n = 0, \pm 1, \pm 2,... .$
Accordingly there is an infinite number of bound states related to a point spectrum of energy values. It extends from $-\infty$ to zero which is a point of accumulation. Because of the open parameters $p_{m, 0}$ there appears a not specified energy scale. As for scattering we have therefore the problem to fix the parameters $p_{m, 0}$ in discussing models which lead to Eq. (\ref{-Re}) in some limiting case.

\bigskip

{\bf 7. Conclusions}

\medskip

We have considered quantum mechanical elastic scattering in the presence of two cylindrically symmetric  electric and magnetic potentials which are singular for $\rho=0.$ The approach was based on a decomposition with regard to partial waves with angular quantum number $m.$ The underlying  Hamiltonian has been made self-adjoint by a self-adjoint extension which fixed the form of the Schr\"{o}dinger solutions for given $m.$ It thereby turned out that for fixed $m$ and given particle momentum $p$ the wave functions belonging to the self-adjoint extension contain open parameters. This is the case for quantum numbers $m$ with $\nu^2:=(m-\beta)^2-\gamma^2 < 1$ where $\beta$ and $\gamma$ are the magnetic-field and charge parameters. To specify the quantum states it is therefore not enough to fix the strength of the singular potentials by these physical parameters. This goes back to the fact that many different physical situations with non-singular electric and magnetic potentials (for example, $\rho=0$ may be excluded for some reason) can lead in some limit to the radial differential equation treated above. Dependent on the situation, the open parameters  may thereby be fixed. It has not been answered above of what type these physical situations are. We showed that a cylinder with hard core is not a representative. 

Independent of the answer to this question one can study the elastic scattering in these singular potentials. We have shown how the AB contribution dominates the behaviour for small scattering angles. The dependence on the charge parameter $\gamma$ and the momentum $p$ is discussed.

There is a lesson to be learned: The open parameters appear mode by mode. The angular momentum quantum number $m$ is essentially an impact parameter. One could think of avoiding the singularities in modifying the physical situation for certain modes (for example by introducing absorption) and keeping other modes unaltered. It may then be that for these unaltered modes one is again confronted with the problem of open parameters which are to be fixed by some physical argument.

\bigskip

{\bf Appendix}

\medskip 

{\bf A: Self-adjoint extension approach}

\medskip

In this Appendix we  present very briefly the ideas and results concerning the self-adjoint extension approach to the problem under consideration. We do not dwell on the rigorous formulations and details. A more extended version will be presented elsewhere.

The $H$ of Eq. (3) is only a formal differential expression. It must be supplemented by specifying the dense domain ${\cal D}_H \subset {\cal H} = {\cal L}_2(R_3),\, {\overline{\cal D}}_H = {\cal H}$ on which $H$ of Eq. (3) defines a self-adjoint operator: $H=H^\dagger.$ This is the problem to be solved. In contrast to the case of regular potentials, in this case the differential expression (3) does not at all define an operator (or even a self-adjoint one) on the natural set of sufficiently smooth functions. The reason is the strong singularity of the potentials (1) and (2) on the $z$-axis.

The differential expression (3) naturally defines an operator $H_0$ on the domain ${\cal D}_{H_0}$ of functions $\psi_0$ which are sufficiently smooth and sufficiently rapidly vanishing when approaching the $z$-axis (and infinity). The operator $H_0$ is evidently {\sl symmetric} (hermitian)\footnote{In fact the equality $(\psi_0^{(1)}, H_0\psi_0^{(2)}) = (H_0\psi_0^{(1)}, \psi_0^{(2)}),\, \psi_0^{(1)}, \psi_0^{(2)} \in {\cal D}_{H_{0}}$ holds true
because the integration by parts is possible without any boundary terms at the $z$-axis (and at infinity).} but it is {\sl not self-adjoint}: $H_0\subset H_0^\dagger.$ The adjoint operator $H_0^\dagger$ is defined by the same differential expression (3) but on the different domain ${\cal D}_{H_0^\dagger} \subset {\cal H},$ which is the extension of ${\cal D}_{H_0}$ with ${\cal D}_{H_0} \subset {\cal D}_{H_0^\dagger}.$ The functions $\psi$ of ${\cal D}_{H_0^\dagger}$ may have singularities on the $z$-axis.\footnote{The integration by parts in $(\psi, H_0\psi_0)$ is possible without boundary terms because of the vanishing of $\psi_0$ on the $z$-axis to give $(H_0^\dagger\psi, \psi_0).$} 

The physically meaningful self-adjoint Hamiltonian $H=H^\dagger$ can be defined as a self-adjoint extension of the symmetric operator $H_0$ (and consequently a restriction of $H_0^\dagger$): $H_0\subset H \subset H_0^\dagger$ and therefore ${\cal D}_{H_0}\subset {\cal D}_{H}\subset {\cal D}_{H_0^\dagger}.$

Now we recall the basic points of the self-adjoint extension theory \cite{Reed75}. For an arbitrary symmetric operator $H_0\subseteq H_0^\dagger$ the possibility to have a self-adjoint extension depends on the so called deficiency subspaces ${\cal H_+}$ and ${\cal H_-}$ which are the eigensubspaces of the operator
$H_0^\dagger$ corresponding to the respective eigenvalues $+i$ and
$-i$\footnote{$\pm i$ can be replaced by any complex numbers $\zeta,\,
\overline\zeta,\, \Im\zeta >0$)}: ${\cal H_\pm} = \left\{\psi_\pm: \psi_\pm
\in {\cal D}_{H_0^\dagger},\, H_0^\dagger\psi_\pm = \pm i\,\psi_\pm\right\}.$
The extension does exist if the dimensions of these subspaces, the so called deficiency indices $n_\pm = \dim{\cal H_\pm}$, coincide: $n_+=n_-.$

If ${\cal H_\pm}$ is trivial, i.e. for $n_+=n_-=0,$ the self-adjoint extension is unique. It is the closure of $H_0: H=\overline H_0=H_0^\dagger.$ In this case the symmetric $H_0$ is called essentially self-adjoint . If ${\cal H_\pm}$ is nontrivial, i.e. for $n_+=n_-=n>0,$ the self-adjoint extension is non-unique. The concrete self-adjoint extension $H_U$ is specified by a unitary mapping $U:{\cal H}_+ \rightarrow {\cal H}_-$ with ${\cal D}_{H_U} = \left\{\psi_U:\psi_U=\psi_0+\psi_+ + U\psi_+,\,\forall \psi_0\in {\cal D}_{H_0},\,\forall \psi_+\in {\cal H}_+; U\psi_+\in {\cal H}_-\right\}.$ $H_U$ coincides with $H_0^\dagger$ on ${\cal D}_{H_U}: H_U\psi_U= H_0\psi_0+i\,\psi_+ -i\,U\psi_+.$\footnote{If $\{\psi_{\pm k}\} with k=1, 2,...,n$ are respective
orthonormal basises in ${\cal H}_\pm,$ then $U$ and ${\cal H}_U$ are specified by a unitary matrix $U_{k l}: {\cal D}_{H_U} = \left\{\psi_U:\, \psi_U=\psi_0+\sum_{k=1}^n c_k\psi_{+, k}+\sum_{k, l=1}^n c_k U_{k l}\psi_{-,
l}\right\},\, H_U\psi_U = H_0\psi_0 + i\sum_{k=1}^n c_k\psi_{+, k}-i\sum_{k,
l=1}^n c_k U_{k l}\psi_{-, l}.$} Therefore if $n_+=n_-=n>0$ there is a manifold of self-adjoint extensions of the given symmetric operator $H_0$ which is the group $U(n).$ In this case additional considerations are necessary to select a particular extension.

In our case we have $n_+=n_-=\infty.$ It is easy to verify that the functions
$$
\psi_\pm = e^{im\varphi}\int_{-\infty}^{\infty}dk\,e^{ikz}\tilde{\psi}_\pm (k)\, H^{(1)}_{\nu}(\sqrt{\pm 2iM -k^2}\rho)
$$
are normalizable solutions of the respective differential equations $H_0^\dagger\psi_\pm = \pm i\psi_\pm$ where $\tilde\psi_\pm (k)$ are arbitrary functions with a compact support. $\nu^2\equiv \nu^2 (m) = (m-\beta)^2- \gamma^2<1$ and may be negative, $m=0, \pm 1, ...,$ $H^{(1)}_{\nu}$ is the Hankel function and $\Im\sqrt{\pm 2iM -k^2}>0.$\footnote{It is also evident that if $\psi_+$ satisfies the equation $H_0^\dagger \psi_+ = +i\,\psi_+,$ then $\psi_- = \overline\psi_+$ satisfies the equation $H_0^\dagger \psi_- = -i\,\psi_-.$} Therefore there exists an infinite-dimensional $U(\infty)$ manifold  of self-adjoint extensions $H_U$ of the symmetric operator $H_0$ in question.

In order to specify further concrete extensions we impose as an additional demand that $H_U$ is invariant under translations along and rotations around the $z$-axis as has been done in the sections above. $H_0$ and $H_0^\dagger$ show this invariance but $H_U$ does not, in general. The reason is that although $H_0$ and $H_0^\dagger$ are invariant under these transformations, the mapping $U$ may not respect this invariance. Such a possibility corresponds to the case where the potential has additional singularities of the $\delta(z)$ -type that brake the translation and rotation invariance. The operator $H_0$ remains the same.

The requirement of this invariance greatly restricts the arbitrariness in
the choice of $U$ but in fact does still not eliminate it. This is in agreement 
with the results of the ``pragmatic approach'' of Sect. 3. This can be 
demonstrated as follows: If we require the rotation and translation 
invariance with respect to the $z$-axis, then the problem of a self-adjoint 
extension of $H_0$ reduces to the problem of self-adjoint extensions of the 
radial symmetric operators
\begin{equation}\label{radH}
H_{m0} = - {d^2\over d\rho^2} -{1\over\rho}{d\over d\rho} + {\nu^2\over\rho^2}\,,
\end{equation}
in the radial Hilbert space ${\cal L}_2 (R_+, \rho d\rho)$ for each given quantum number $m=0,\pm 1, ... .$ These operators are defined on functions $R_{m0}(\rho)$ which vanish sufficiently rapid for $\rho\rightarrow
0.$ The deficiency indices of $H_{m0}$ are fitted by $\nu^2=\nu^2(m)$. It is
easy to see that if $\nu^2\geq 1$ we have $n_+=n_-=0.$ Consequently $H_{m0}$ is essentially self-adjoint and there is no arbitrariness in defining the
self-adjoint extension $H_{m}.$ But if $\nu^2 < 1,$ we have $n_+=n_-=1,$ and for each $m$ we have a 1-parameter $U(1)$ manifold of self-adjoint
extensions $H_{m}.$ We can specify the domains ${\cal D}_{H_m}$ of these
extensions in terms of the boundary behaviour of the functions $R_m (\rho)$ for $\rho\rightarrow 0.$ For $\nu^2\geq 1$ we have $R_m (0)=0.$ If $0<\nu^2 < 1 $\footnote{The case of $\nu=0$ requires a special consideration.} a concrete self-adjoint extension 
$H_{m l_m}$ is specified by the (asymptotic) boundary condition of the form
\begin{equation}\label{asybc1}
R_m (\rho):=R_{m l_m}(\rho) \sim \rho^{\mu} + l_m \rho^{-\mu},\quad \rho\rightarrow 0
\end{equation}
with a particular parameters $l_m \in R_1.$ If $\nu^2 < 0,$ the concrete
self-adjoint extension $H_{m \vartheta_m}$ is defined by the (asymptotic)
boundary condition of the form
\begin{equation}\label{asybc2}
R_m (\rho):=R_{m \vartheta_m}(\rho) \sim \rho^{i\mu} + e^{i\vartheta_m} \rho^{-i\mu},\quad \rho\rightarrow 0
\end{equation}
with the particular parameters $\vartheta_m,\, 0\leq \vartheta_m<2\pi.$ The same behaviour has been obtained in Sect. 3.

\medskip

{\bf To sum up:} We can determine the eigenfunctions of the self-adjoint Hamiltonian for the problem under consideration by separating variables in cylindrical coordinates and imposing the boundary conditions stated above on the radial wave functions at $\rho=0$ in fixing $\lambda_m$ and $\theta_m.$ This specifies the Hamiltonian as a self-adjoint operator.

The arbitrariness in defining the Hamiltonian is finite-parametric. The
number of parameters agrees with the number of the orbital quantum numbers
$m=0, \pm 1,...,$ for which $\nu^2=(m-\beta)^2-\gamma^2 < 1.$ The further
reduction of this finite-parametric arbitrariness requires physical arguments in addition (to the rotation and translation invariance).

\bigskip

{\bf B: Some integrals with the Bessel functions}

\medskip

Using formula 5.53 of \cite{Gradshteyn80} we obtain:
\begin{eqnarray} \label{i1}
&&\int_0^{\infty} J_\nu (p'\rho) J_\nu(p\rho)\rho d\rho =  \\
&&\lim_{R\rightarrow\infty} {1\over p^2-p'^2}
\left[p'\rho J_\nu(p\rho) J_{\nu-1}(p'\rho) -
p\rho J_{\nu-1}(p\rho) J_\nu(p'\rho)\right]\big|^R_0 =\nonumber \\
&&= \lim_{R\rightarrow\infty} {2\over \pi(p^2-p'^2)}
\left[\sqrt{p'\over p}\cos(pR-{\pi\over2}\nu-{\pi\over 4})
\cos(p'R-{\pi\over 2}(\nu-1)-{\pi\over 4})\right.- \nonumber \\
&&\left.\sqrt{p\over p'}\cos(pR-{\pi\over 2}(\nu-1)-{\pi\over 4})
\cos(p'R-{\pi\over 2}\nu-{\pi\over 4})\right] = \nonumber \\
&&{1\over \pi\sqrt{pp'}}\lim_{R\rightarrow\infty}\left[{\sin(p-p')R\over p-p'} - {\cos(p+p')R\over p+p'}\cos\pi p - {\sin(p+p')R\over p+p'}\sin\pi p\right] = \nonumber \\
&&{\delta(p-p')\over\sqrt{pp'}} \nonumber
\end{eqnarray}
and
\begin{eqnarray} \label{i2}
&&\int_0^{\infty} J_{-\nu} (p'\rho) J_\nu(p\rho)\rho d\rho = \\
&&\lim_{R\rightarrow\infty} {1\over p^2-p'^2}
\left[p'\rho J_\nu(p\rho) J_{-\nu-1}(p'\rho) -
p\rho J_{\nu-1}(p\rho) J_{-\nu}(p'\rho)+2\nu J_\nu(p\rho) J_{-\nu}(p'\rho)\right]\big|^R_0 \nonumber \\
&&= \lim_{R\rightarrow\infty} {2\over \pi(p^2-p'^2)}
\left[\sqrt{p'\over p}\cos(pR-{\pi\over2}\nu-{\pi\over 4})
\cos(p'R+{\pi\over 2}(\nu+1)-{\pi\over 4})\right.- \nonumber \\
&&\left.\sqrt{p\over p'}\cos(pR-{\pi\over 2}(\nu-1)-{\pi\over 4})
\cos(p'R+{\pi\over 2}\nu-{\pi\over 4})\right] - \nonumber \\
&&{2\over p^2-p'^2}\left({p\over p'}\right)^\nu \left[{1\over\Gamma(\nu+1)\Gamma(-\nu)} - {1\over\Gamma(\nu)\Gamma(-\nu+1)} +
{2\nu\over\Gamma(\nu+1)\Gamma(-\nu+1)}\right] = \nonumber \\
&&{1\over\pi\sqrt{p p'}}\lim_{R\rightarrow\infty}
\left[{\sin[(p-p')R-\pi\nu]\over p-p'} - {\cos(p+p')\over p+p'}\right] - \nonumber \\
&&{2\over (p^2-p'^2)}\left({p\over p'}\right)^\nu {1\over \Gamma(\nu+1)\Gamma(-\nu)} =
{\delta(p-p')\over\sqrt{p p'}}\cos\pi\nu + {2\sin\pi\nu\over\pi(p^2-p'^2)}
\left({p\over p'}\right)^\nu\,. \nonumber
\end{eqnarray}
We have thereby made use of 
\begin{eqnarray} \label{d+}
&&\int_0^\infty e^{ixt} dt = \pi\delta_+(x)=\pi\delta(x)+iP\left({1\over x}\right) = \nonumber \\
&&\lim_{R\rightarrow\infty}\int_0^R e^{ixt} dt = \lim_{R\rightarrow\infty}
\left({\sin xR\over x} + i{1-\cos xR\over x}\right)\,,
\end{eqnarray}
so that we may write
\begin{equation}
\delta(x) = \lim_{R\rightarrow\infty}{\sin xR\over \pi x}, \quad P\left({1\over x}\right)=\lim_{R\rightarrow\infty}{1-\cos xR\over x}\,.
\end{equation}

\bigskip

\section*{Acknowledgments}

V.~S.~thanks Professor J.~Audretsch and the members of his group for the
friendly atmosphere and the hospitality at the University of Konstanz. This
work was supported by the Deutsche Forschungsgemeinschaft. B.~V. is grateful
for the support from RFBR (Grant No. 99-01-00376).


\end{document}